\documentclass[runningheads]{llncs}
\usepackage[T1]{fontenc}
\usepackage{graphicx}
\usepackage{multirow}
\usepackage{amsmath}
\usepackage{booktabs,subcaption,amsfonts,dcolumn}
\newcolumntype{d}[1]{D..{#1}}
\usepackage{xcolor,colortbl}
\usepackage[misc,geometry]{ifsym} 
\definecolor{Gray}{gray}{0.85}

\newcolumntype{a}{>{\columncolor{Gray}}c}
\begin{document}
\title{Opinions Vary? Diagnosis First!}
\author{Junde Wu\and
Huihui Fang \and
Dalu Yang \and Zhaowei Wang \and Wenshuo Zhou \and Fangxin Shang \and Yehui Yang\and Yanwu Xu\textsuperscript{ (\Letter)}}

\authorrunning{J. Wu et al.}

\institute{Healthcare Group, Baidu\\
\email{ywxu@ieee.org}
\\}

\maketitle              
\begin{abstract}
With the advancement of deep learning techniques, an increasing number of methods have been proposed for optic disc and cup  (OD/OC) segmentation from the fundus images. Clinically, OD/OC segmentation is often annotated by multiple clinical experts to mitigate the personal bias. However, it is hard to train the automated deep learning models on multiple labels. A common practice to tackle the issue is majority vote, e.g., taking the average of multiple labels. However such a strategy ignores the different expertness of medical experts. Motivated by the observation that OD/OC segmentation is often used for the glaucoma diagnosis clinically, in this paper, we propose a novel strategy to fuse the multi-rater OD/OC segmentation labels via the glaucoma diagnosis performance. Specifically, we assess the expertness of each rater through an attentive glaucoma diagnosis network. For each rater, its contribution for the diagnosis will be reflected as an expertness map. To ensure the expertness maps are general for different glaucoma diagnosis models, we further propose an Expertness Generator (ExpG) to eliminate the high-frequency components in the optimization process. Based on the obtained expertness maps, the multi-rater labels can be fused as a single ground-truth which we dubbed as Diagnosis First Ground-truth ($DiagFirst$GT). Experimental results show that by using $DiagFirst$GT as ground-truth, OD/OC segmentation networks will predict the masks with superior glaucoma diagnosis performance.

\end{abstract}
\section{Introduction}\label{sec:introduction}
Accurate optic disc and cup (OD/OC) segmentation on fundus images is important for the clinical assessment of glaucoma. It has been increasingly popular to develop automated OD/OC segmentation \cite{mark1,ji2021learning}ß, which is especially accelerated by the exciting breakthroughs of deep neural networks and publicly available datasets \cite{seg_data,fang2022refuge2,wu2022gamma}. Different from the natural image dataset, OD/OC segmentation datasets are often annotated by multiple clinical experts (e.g., RIGA \cite{seg_data}, REFUGE-2 \cite{fang2022refuge2}) to mitigate the subjective bias of particular rater. However, this multi-rater scenario also brings challenges to the deployment of deep learning models, since they are commonly trained by a single ground-truth. For training the neural network in multi-rater scenario, a common practice is to take the average of multiple labels, i.e., majority vote. Although this fusion strategy is simple and easy to implement, it comes at the cost of ignoring the different expertness of multiple experts \cite{fu2020retrospective}. 

It is thus necessary to estimate the rater expertness for the fusion of multi-rater annotations. A popular way is to estimate the expertness based on the raw image prior \cite{cao2019max,wu2022learning,raykar2010learning}. However, we note that OD/OC segmentation is clinically established for the glaucoma diagnosis. Specifically, the Cup-to-Disc Ratio (CDR) calculated from the segmentation masks is an essential clinical parameter for the glaucoma diagnosis  \cite{garway1998vertical}. This inspires us to take glaucoma diagnosis as a gold standard to estimate the multi-rater expertness and fuse the multi-rater labels. The fused label then can be used as the ground-truth for the segmentation training.
 
Specifically, we use a glaucoma diagnosis network to evaluate the multi-rater expertness. The evaluation diagnosis network is implemented by a segmentation attentive diagnosis network, following  \cite{zhou2019collaborative}. We optimize an expertness map for each rater to maximize the glaucoma diagnosis performance of the network. The ground-truth fused according to this multi-rater expertness maps is named as Diagnosis First Ground-truth ($DiagFirst$GT). In order to make $DiagFirst$GT general for different diagnosis networks, we further propose Expertness Generator (ExpG) to generate the expertness maps. ExpG helps to constrain the high-frequency components by the continuity nature of neural network, so as to promise the effectiveness of $DiagFirst$GT across different diagnosis networks.

In brief, three major contributions are made with this paper. 
First, we propose a novel strategy to fuse the ground-truth from multiple OD/OC segmentation labels. The obtained ground-truth is called $DiagFirst$GT, which can significantly facilitate the glaucoma diagnosis. Secondly, in order to reinforce the generalization of $DiagFirst$GT, we further propose ExpG to generate the expertness maps in the optimization process. ExpG eliminates the high-frequency components by the continuity of neural network. Finally, the experimental results show the OD/OC segmentation models trained on $DiagFirst$GT estimates the masks with a clear 3\% AUC improvement on standard glaucoma diagnosis network. Such a method can be freely applied to any model architecture and gains general improvement without external data.

\section{Methodology}
\subsection{Motivation}
In clinical research, a better segmentation often lead to a better diagnosis. In the neural network models, many prior work \cite{zhou2019collaborative,as2,wu2020leveraging} also show that lesions segmentation masks can bring diagnosis networks with solid improvement. Move a step further, a natural question is that whether the segmentation masks with better qualities are better for the disease diagnosis? 
\begin{figure}[h]
    \centering
    \includegraphics[width=0.8\textwidth]{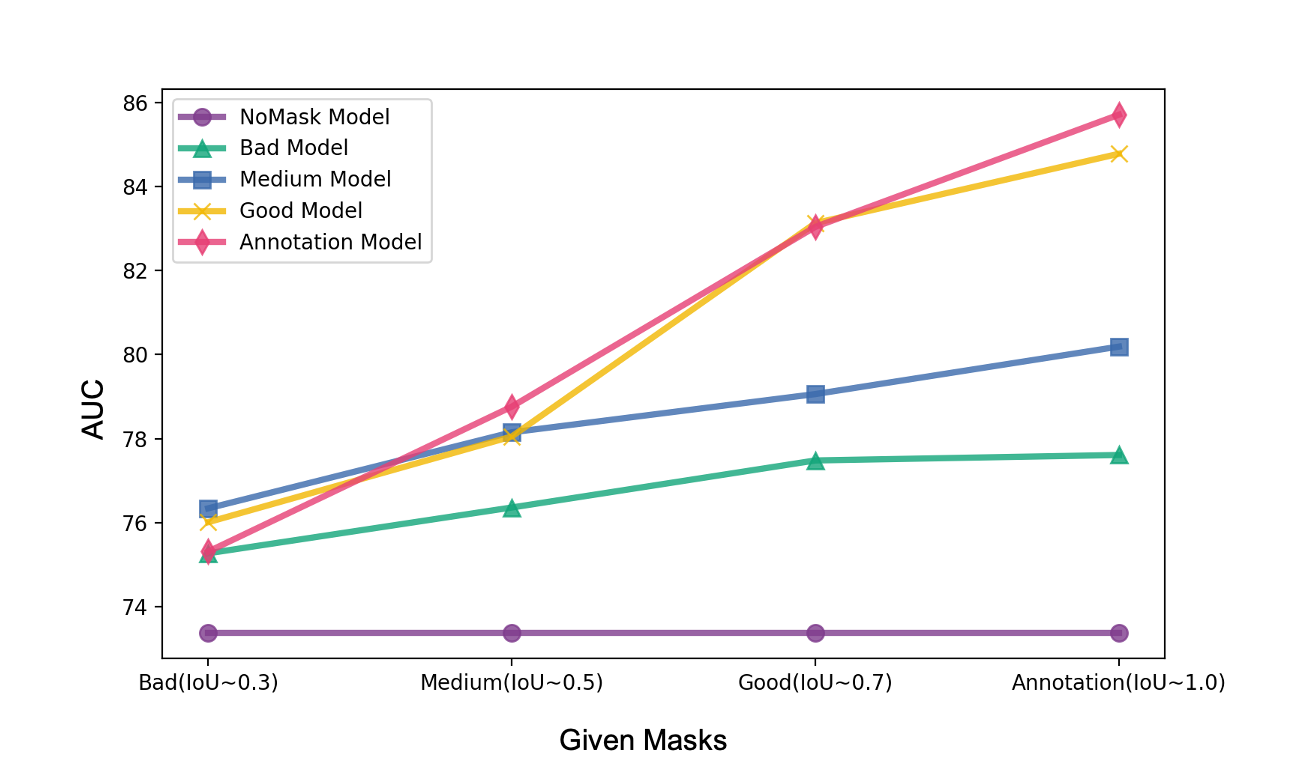}
    \caption{Diagnosis performance of segmentation masks with different qualities. It shows the OD/OC mask quality and glaucoma diagnosis performance are positively correlated.}
    \label{fig:motivation}

\end{figure}

We design a simple experiment on fundus images to answer this question. We train ResNet50 \cite{resnet} to diagnose glaucoma from the fundus images, with OD/OC segmentation masks as the auxiliary information. The fundus image and its segmentation mask are concatenated as the input of the network. The network is supervised by the glaucoma classification labels on a private dataset with 1600 samples. Segmentation masks with different qualities are collected from early-stop training, which are denoted as Bad (average IOU of segmentation is 0.3), Medium (= 0.5), Good (= 0.7), and Annotation. We use these masks to train the corresponding diagnosis networks, denoted as Bad Model, Medium Model, Good Model, and Annotation Model. Then the segmentation masks with different qualities are given to each model for the inference. The final diagnosis performance measured by AUC score (\%) is shown in Figure \ref{fig:motivation}. In the figure, NoMask Model denotes training the diagnosis network only on the raw images.

It is clear to see that better segmentation masks promote the diagnosis performance more than the worse. The diagnosis networks trained on better segmentation masks show better average diagnosis performance. In the inference, the diagnosis performance is also gradually improved with the improvement of the segmentation quality. This result suggests, like what in clinical research, better OD/OC segmentation is also more conducive to the glaucoma diagnosis in the neural network models.

\subsection{Learning $DiagFirst$GT}
\begin{figure*}[!t]
\centering
\includegraphics[width=1\textwidth]{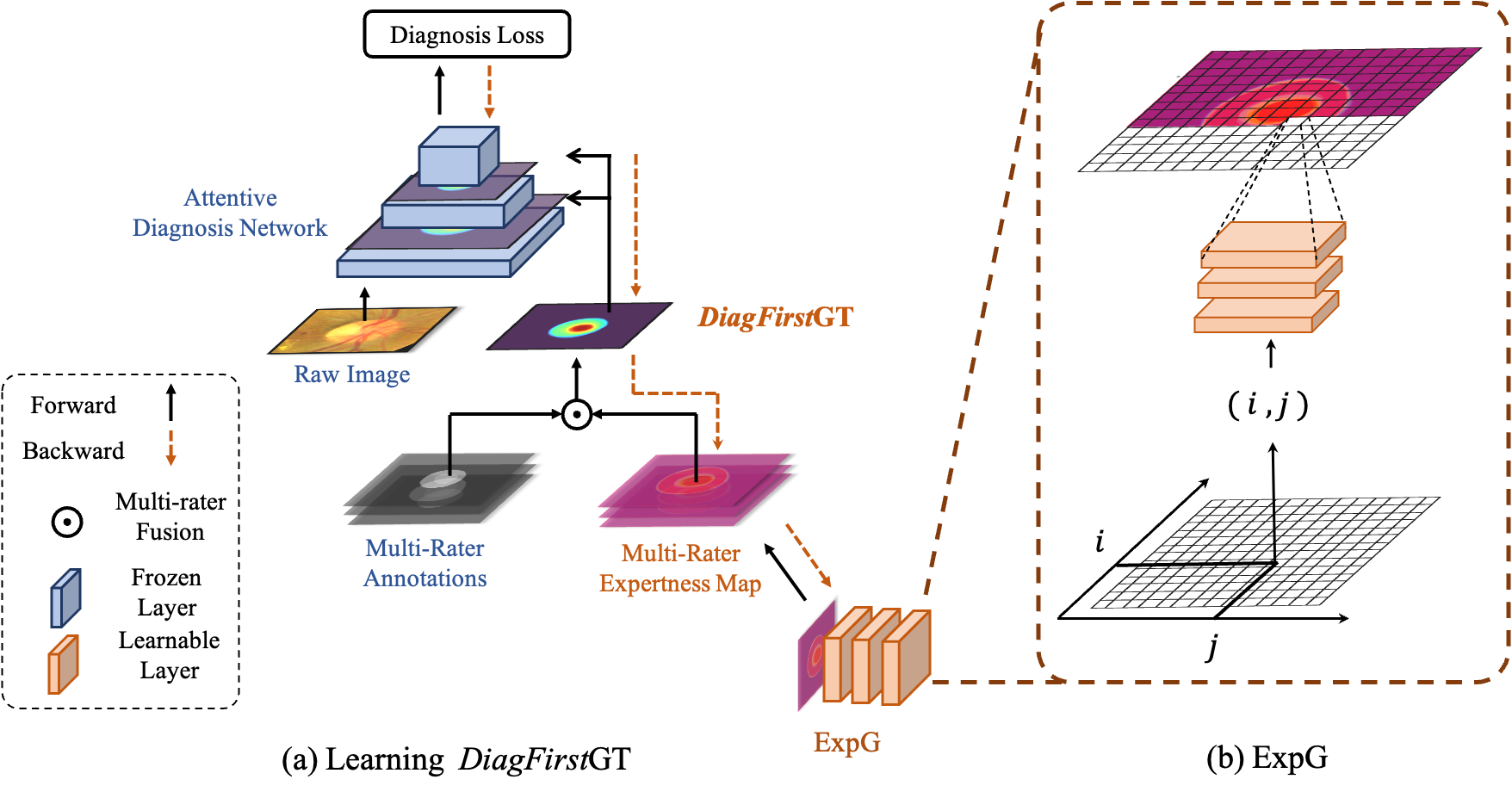}
\caption{Overall flow of proposed method. Blue denotes static parameters. Orange denotes learnable parameters.}
\label{fig:overall}
\end{figure*}
Motivated by this relationship between OD/OC segmentation and glaucoma diagnosis, we want to find the optimal OD/OC segmentation label toward the glaucoma diagnosis when multiple labels are collected. Such a label, based on the results above, may be closer to the potential gold segmentation. To fuse multiple labels to one ground-truth, we weighted sum the multiple labels by the expertness of each rater:
\begin{equation}
   \textrm{Groundtruth} = s \odot m  = \sum_{i=1}^{n} s_{i} * m_{i}, 
\end{equation}\label{eqn:fusion}
where $*$ denotes the element-wise multiplication, $s_{i}$ and $m_{i}$ are the annotation and the expertness map of the rater $i$, respectively.

We then assess the rater expertness through the glaucoma diagnosis. In other words, the rater contributes more to the correct diagnosis would be given higher expertness. Toward that end, we take the multi-rater expertness maps as the learnable variables to maximize the diagnosis performance of a OD/OC attentive diagnosis network. The network is pre-trained on the same data so as to correctly evaluate the diagnosis contribution. The optimization is solved by the gradient descent to activate the target class in diagnosis network. We dubbed the label fused as $DiagFirst$GT. 

The overflow of learning $DiagFirst$GT is shown in Figure \ref{fig:overall} (a). Formally, consider each raw image $x \in \mathbb{R}^{h\times w\times c}$ is annotated by $n$ raters, resulting in $n$ segmentation masks $s \in \mathbb{R}^{h\times w \times n}$, and the image is diagnosed as $y \in [0,1]$. Let $\theta$ denotes the set of diagnosis model parameters. $L (\theta, x, y)$ denotes the loss function of a standard classification task. Our goal is to find an optimal expertness map $m \in \mathbb{R}^{h\times w\times n}$ by solving the following optimization problem:
\begin{equation}\label{first order backer}
    m^{*} = \mathop{\arg\min}\limits_{m} L (\theta,x \oplus [s \odot m] ,y), 
\end{equation}
where $\oplus$ denotes concatenation operation, $m^{*}$ denotes the optimal expertness maps. 
Note that we applied Softmax on $m$ to normalize the weights. According to Eqn. (\ref{first order backer}) , we are actually finding the expertness maps which can minimize the diagnosis loss. We name the fused ground-truth under these expertness maps $s \odot m^{*}$ as $DiagFirst$GT.

Gradient descent is then adopted to solve Eqn. (\ref{first order backer})
\begin{equation}\label{gradient}
    m^{t+1} = m^{t} + \alpha \nabla_{m} L (\theta,x \oplus [s \odot m] ,y),
\end{equation}
where $\alpha$ is the learning rate. The fusion ground-truth optimized in this way can improve the diagnosis performance to a very high level, but since it is optimized toward one specific diagnosis network, it is not general enough. The visualization results also show it suffers heavily from high-frequency noises. A visualized example is shown in Figure \ref{fig:supvis}. The latest findings suggest that these high-frequency components have close relationship with the
generalization capability \cite{Wu2019UniversalTA,h3}. Specifically, high-frequency components are generated by the repeated grid effect of the transposed convolution  \cite{warfield2004simultaneous} when we backprop gradients through each convolution layer. Thus they are very specific to the network architecture, like the number of the layers, the stride of the convolution, etc. 

\subsection{ExpG}
Therefore, a possible approach to improve generalization of $DiagFirst$GT is to constrain the high-frequency components in the optimization process. We tried several methods to achieve the goal, including Transformation Robustness (TransRob), Fourier Transform (Fourier) and proposed Expertness Generator (ExpG). Transformation Robustness constrains high-frequency gradients by applying small transformations to the expertness map before optimization. In practice, we rotate, scale and jitter the maps. Fourier Transform transforms the expertness map parameters to frequency domain, thus decorrelated the relationship between the neighbour pixels. 

ExpG constrains the high-frequency components by the continuity nature of the neural network, which is implemented by a tiny CNN based pixel generator. An illustration of ExpG is shown in Figure \ref{fig:overall} (b). The network scans one pixel at a time. For each pixel it predicts the pixel value given the position of the pixel. The input of ExpG is the coordinate vector $ (i,j)$ and the output is the pixel value. ExpG is optimized to generate expertness maps which are able to minimize the diagnosis loss. Denote the parameters of ExpG as $\phi$, our goal is to solve:
\begin{equation}\label{eqn:lhc}
    \mathop{\arg\min}\limits_{\phi} L (\theta,x \oplus [s \odot \{\rm{ExpG}_{\phi} (i,j)\}_{i = 1 \sim h}^{j = 1 \sim w}] ,y),
\end{equation}
where $\{\rm{ExpG}_{\phi} (i,j)\}_{i = 1 \sim h}^{j = 1 \sim w}$ denotes generated expertness map with size $h \times w \times n$ by ExpG. Since the continuity of the neural network mapping function, similar inputs tend to cause similar outputs, which lead the element values in expertness maps variant smoothly between the positions and thus eliminate the high-frequency components. 

\begin{figure}[h]
    \centering
    \includegraphics[width=0.8\textwidth]{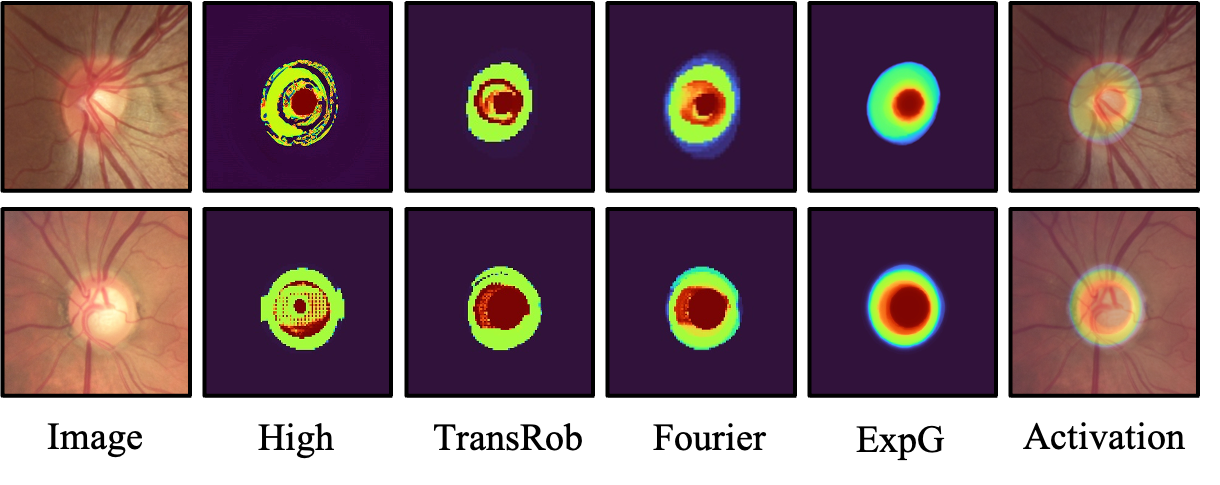}
    \caption{Visualized results of various high-frequency elimination methods and ExpG. From left to right are raw fundus image, original high-frequency results, TransRob, Fourier, ExpG low-frequency results and activation results of ExpG.}
    \label{fig:supvis}

\end{figure}

The visualized results of all high-frequency elimination methods are shown in Figure \ref{fig:supvis}. The first line is a non-glaucoma example and the second line is a glaucoma example. We can see the visualized effects of TransRob and Fourier are obviously improved comparing with high-frequency examples. But ExpG achieves much better visualized effects than both of them. Thus, we adopt ExpG to eliminate the high-frequency components in the experiments. 

\section{Experiment}
\subsection{Experimental settings}
All the experiments are implemented with the PyTorch platform and trained/tested on 4 Tesla P40 GPU with 24GB of memory. All images are uniformly resized to the dimension of 256$\times$256 pixels. The networks are trained in an end-to-end manner using Adam optimizer with a mini-batch of 16 for 80 epochs. The learning rate is initially set to 1 $\times 10^{-4}$. The detailed hyper-parameters and architectures can be found in the code.

We conduct the experiments on REFUGE-2 \cite{fang2022refuge2} dataset. REFUGE-2 is a publicly available dataset for retinal cup and disc segmentation annotated by multiple glaucoma experts. REFUGE-2 contains in total 2000 color fundus images, including three sets each with 1200 images for training, 400 for validation, and 400 for testing. Seven glaucoma experts from different organizations labeled the optic cup and disc contour masks manually for the REFUGE-2  benchmark. 200 samples correspond to glaucomatous subjects, and the others correspond to non-glaucomatous subjects. The glaucomatous subjects are distributed equally to the training, validation, and test set. It is worth noting that the diagnosis network is pre-trained on REFUGE-2 training set to generate $DiagFirst$GTs, thus no external data is used.

\subsection{Comparing with SOTA}
To verify the glaucoma diagnosis effect of $DiagFirst$GT, we compare it with state-of-the-art (SOTA) multi-rater fusion strategies on REFUGE-2 dataset. Specifically, we train standard UNet on various multi-rater fusion ground-truths, including random fusion (Random), majority vote (MV), STAPLE  \cite{warfield2004simultaneous}, AggNet  \cite{albarqouni2016aggnet} and Max-Mig \cite{cao2019max}. We then evaluate the diagnosis performance of their estimated masks on the test set. The glaucoma diagnosis performance is evaluated by a range of OD/OC segmentation-assisted glaucoma prediction methods, including which based on vertical Cup-to-Disc Ratio (vCDR)  \cite{chandrika2013analysis,thangaraj2017glaucoma} and which based on deep learning models  \cite{li2019attention,luo2021segmentation,wu2020leveraging,zhang2019attention}. The obtained diagnosis performance measured by AUC score (\%) is shown in Table \ref{Table:GoldDiag}.

\begin{table}[h]
\setlength\arrayrulewidth{0.1pt}
\caption{Comparison of $DiagFirst$GT and SOTA multi-rater fused ground-truths on a range of diagnosis models (a) and segmentation models (b).}
\begin{subtable}[t]{0.48\textwidth}
\flushleft
\resizebox{0.99\columnwidth}{!}{%
\begin{tabular}{c|cccccc}
\toprule

\multicolumn{1}{c|}{\multirow{3}{*}{\begin{tabular}[c]{@{}c@{}} Ground\\ Truth\end{tabular}}} & \multicolumn{6}{c}{Diagnosis Models}                                                                                                                                                                                   \\ \cline{2-7}
\multicolumn{1}{c|}{} &  \multicolumn{2}{c}{vCDR} & \multicolumn{4}{c}{Deep Learning} \\
\cline{2-7}
\multicolumn{1}{c|}{}                                                                                &  \cite{chandrika2013analysis}           &  \cite{thangaraj2017glaucoma}           & \multicolumn{1}{c}{ \cite{luo2021segmentation}} & \multicolumn{1}{c}{ \cite{li2019attention}} & \multicolumn{1}{c}{ \cite{zhang2019attention}}  & \multicolumn{1}{c}{ \cite{wu2020leveraging}} \\ \hline
NoMask   & -       & -     & 77.45   & 79.21    & 80.06    & 79.18                                \\
Random     & 68.79      & 73.34       & 79.74     & 82.76    & 84.34    & 85.07                                \\
MV  & 72.20  & 76.04  & 80.54     & 83.65    & 85.55   & 86.23                                \\
STAPLE     & 71.66   & 77.41    & 81.23    & 84.17    & 86.18     & 87.00                                \\
AggNet       & 71.25  & 75.91 & 80.42  & 83.80  & 85.75  & 87.50                                \\
Max-Mig     & 70.14  & 75.02  & 80.31  & 83.88 & 85.73   & 86.76  \\ \hline
$DiagFirst$GT    & \textbf{74.12} & \textbf{79.23} & \textbf{82.20}      & \textbf{86.27}     & \textbf{88.86} & \textbf{89.62}      \\ \bottomrule
\end{tabular}%
}
\caption{\footnotesize Comparison of $DiagFirst$GT and SOTA multi-label fusion methods. Taking $DiagFirst$GT as the ground-truth, the segmentation model estimates better masks for glaucoma diagnosis. The results are verified on vCDR and deep learning based diagnosis methods by AUC (\%).}
\label{Table:GoldDiag}
\end{subtable}
\hspace{\fill}
\begin{subtable}[t]{0.48\textwidth}
\resizebox{0.95\columnwidth}{!}{%
\begin{tabular}{c|cca|cca}
\toprule
      & \multicolumn{3}{c|}{MV} & \multicolumn{3}{c}{$DiagFirst$GT} \\ \hline
      &$\mathcal{D}_{disc}$ &$\mathcal{D}_{cup}$ & AUC   &$\mathcal{D}_{disc}$         &$\mathcal{D}_{cup}$         & AUC     \\ \hline
AGNet & 91.30     & \textbf{77.38}     & 80.18    & \textbf{91.65}      & 75.14      & \textbf{81.58}      \\
CENet & \textbf{91.70}  & \textbf{80.34}  & 80.46    & 91.05      & 78.63      & \textbf{83.41}   \\
pOSAL & \textbf{94.52}     & \textbf{82.42}     & 81.53    & 93.64      & 82.32      & \textbf{84.07}   \\
BEAL  & 93.84     & \textbf{83.51}     & 81.66    & \textbf{93.88}      & 81.93      & \textbf{84.35}   \\
ATTNet & \textbf{96.10}     & \textbf{84.91}     &81.75 & 95.78      & 84.65      & \textbf{84.82}   \\ \bottomrule
\end{tabular}%
}
\caption{\footnotesize Comparison of majority vote (MV) and $DiagFirst$GT ground-truth on various segmentation models measured by Dice (\%) and AUC (\%). Although $DiagFirst$GT is more difficult to learn, it supervises the segmentation models to learn more glaucoma diagnosis knowledge.}
\label{tab:seg}
\end{subtable}
\end{table}



We can see $DiagFirst$GT significantly improves the diagnosis performance on all of the methods, no matter it is vCDR based \cite{thangaraj2017glaucoma,chandrika2013analysis}, attentive CNN based  \cite{li2019attention,zhang2019attention,luo2021segmentation} or transfer learning CNN based \cite{wu2020leveraging}. On vCDR based, attentive network based and transfer learning based methods, it outperforms the previous best ground-truths (MV, STAPLE and LFC) by a 1.82\%, 2.68\% and 2.12\% AUC, respectively. This demonstrates $DiagFirst$GT is a general and effective ground-truth for the glaucoma diagnosis algorithms, and also indicates its significant potential value in clinical diagnosis of glaucoma.

To verify $DiagFirst$GT is also an effective ground-truth for various segmentation models. We train SOTA segmentation models on $DiagFirst$GT, and compare it with traditional majority vote ground-truth (MV). The SOTA segmentation models include AGNet  \cite{acnet}, CENet  \cite{cenet}, pOSAL  \cite{posal}, BEAL  \cite{beal} and ATTNet  \cite{ATTNet}. The segmentation performance measured by optic disc Dice score ($\mathcal{D}_{diusc}$) and optic cup Dice score ($\mathcal{D}_{cup}$) is shown in Table \ref{tab:seg}. We can see the segmentation models perform better on MV than $DiagFirst$GT. That is because MV is a simple average of multiple labels which is easy to learn, while the fusion of $DiagFirst$GT is based on the diagnosis prior knowledge that is more difficult to learn. We also report the glaucoma diagnosis performance of their estimated masks measured by AUC (\%). The diagnosis performance is evaluated by  \cite{zhou2019collaborative} pre-trained on a large glaucoma detection dataset (LAG  \cite{li2019attention}). It is shown although the segmentation models perform better on MV, the diagnosis performance of the estimated masks is worse than that of $DiagFirst$GT. With the improvement of the segmentation capabilities, the glaucoma diagnosis AUC on $DiagFirst$GT increase about 3\%, also outperforms the counterpart on MV (about 1\%), indicating the supervision of $DiagFirst$GT can improve the diagnosis performance of estimated masks to a larger extent. 

\subsection{Ablation study on ExpG}
ExpG is proposed in the paper to improve generalization of $DiagFirst$GT. In order to verify the effectiveness of ExpG, we compare the $DiagFirst$GT before and after the application of ExpG on REFUGE-2 dataset. The diagnosis performance measured by AUC (\%) is shown in Table \ref{Table:generalization}. Experiment is conducted on a range of OD/OC-assisted diagnosis models to evaluate the generalization capability. Six different models are selected, including DualStage \cite{bajwa2019two}, DENet \cite{fu2018disc}, AGCNN  \cite{li2019attention}, ColNet \cite{zhou2019collaborative}, L2T-KT \cite{wu2020leveraging}, and Swin Transformer\cite{liu2021swin} with mask-concated input. 

As shown in Table \ref{Table:generalization}, comparing with original high-frequency $DiagFirst$GT, ExpG generated $DiagFirst$GT consistently achieves higher performance on different diagnosis methods, especially on stronger networks. Concretely, ExpG generated $DiagFirst$GT outperforms the counterpart by a 3.95\% AUC on DENet, a 4.93\% AUC on AGCNN, a 6.06\% AUC on Swin-cat and a 7.94\% AUC on L2T-KT. It is obvious that ExpG generated $DiagFirst$GT is more robust for different diagnosis networks.

\begin{table}
\centering
\caption{The ablation study of ExpG. High-frequency $DiagFirst$GT and ExpG generated low-frequency $DiagFirst$GT are compared on different diagnosis backbones. The diagnosis performance is evaluated by AUC (\%). }

\begin{tabular}{c|cccccc}
\hline
\multicolumn{1}{c|}{} & \multicolumn{3}{c}{Original } & \multicolumn{3}{c}{ExpG } \\ \hline
DualStage  \cite{bajwa2019two}& \multicolumn{3}{c}{79.36 }  & \multicolumn{3}{c}{\textbf{80.78}}\\
DENet  \cite{fu2018disc}& \multicolumn{3}{c}{81.21 }  & \multicolumn{3}{c}{\textbf{85.16}}\\
AGCNN  \cite{li2019attention}& \multicolumn{3}{c}{79.35 }  & \multicolumn{3}{c}{\textbf{84.28}}\\
Swin-cat \cite{liu2021swin} & \multicolumn{3}{c}{78.66 }  & \multicolumn{3}{c}{\textbf{84.72}}\\
ColNet  \cite{zhou2019collaborative} & \multicolumn{3}{c}{81.69 }  & \multicolumn{3}{c}{\textbf{87.34}}\\
L2T-KT \cite{wu2020leveraging} & \multicolumn{3}{c}{83.67 }  & \multicolumn{3}{c}{\textbf{91.61}}\\
\hline
\end{tabular}
\label{Table:generalization}
\end{table}


\section{Conclusion}
OD/OC segmentation is often annotated by multiple experts. For training standard deep learning models, the multiple labels need to be fused to one ground-truth. The existed fusion methods often ignore the different expertness of each rater. In the paper, we take glaucoma diagnosis as the gold standard to assess the multi-rater expertness, which enable the fused ground-truth $DiagFirst$GT to facilitate the glaucoma diagnosis. Detailed experiments revealed that by training on $DiagFirst$GT, the OD/OC segmentation networks can produce the masks with high glaucoma diagnosis performance. The results are verified on a variety of glaucoma diagnosis networks to promise the generalization. Since OD/OC segmentation is often used to assist the glaucoma diagnosis clinically, we believe the proposed method is in line with the original intention of OD/OC segmentation task and has great significance for the clinical glaucoma diagnosis. 


\bibliographystyle{splncs04}
\bibliography{egbib}

\end{document}